# AnimalCatcher: 動物の多彩な行動を引き出すデジタルカメラ

塚田 浩二[†1]　沖 真帆[†1]　栗原 一貴[†2]　古館 佑子[†2]

**概要**：動物の写真を撮影することは難しい．動物はカメラに対して特に反応しない上に，言葉で指示を出すこともできないので，音を出したりジェスチャをしたりと，あの手でこの手で気を惹く必要がある．それでも意図通りの動きや表情を捉えることは難しく，粘り強くシャッターチャンスを待つ根気勝負になりがちである．そこで本研究では，撮影中に動物の注意を惹く多様な音を指向性スピーカーで照射することで，意図的に動物のリアクションを引き出すカメラ「AnimalCatcher」を提案する．本論文では，AnimalCatcher のプロトタイプを試作し，動物園での運用を通した撮影事例を紹介し，その有効性を検証すると共に，専門家の意見や倫理的な観点も踏まえて議論する．

# AnimalCatcher: a digital camera to capture various reactions of animals

Koji Tsukada[†1]　Maho Oki[†1]　Kazutaka Kurihara[†2]　Yuko Furudate[†2]

***Abstract:*** People often have difficulty to take pictures of animals, since animals usually do not react with cameras nor understand verbal directions. To solve this problem, we developed a new interaction technique, AnimalCatcher, which can attract animals' attention easily. The AnimalCatcher shoots various sounds using directional speaker to capture various reactions of animals. This paper describes concepts, implementation, and example pictures taken in a zoo.

## 1. はじめに

近年，デジタルカメラの普及に伴い，さまざまな場面でデジタル写真／動画を撮影する機会が増加している．旅行先の風景を撮影したり，家族・友人とのスナップ写真を撮影したり，子供やペットの日常を動画撮影したりとその用途は多岐に渡る．その中でも，特にペットを飼っている人にとっては動物の写真を撮影する機会は多い．そうでない人も，動物園などに訪問した際は動物の写真を撮影することは多いだろう．

しかし，一般に動物を撮影することは，人を撮影するよりも難しい．動物はカメラに対して直接反応しない上に，言葉で指示を出すこともできないので，音を出したりジェスチャをしたりと，あの手でこの手で気を惹く必要がある．それでも結局意図通りの動きや表情を捉えることは難しいので，粘り強くシャッターチャンスを待つ根気勝負になりがちである．

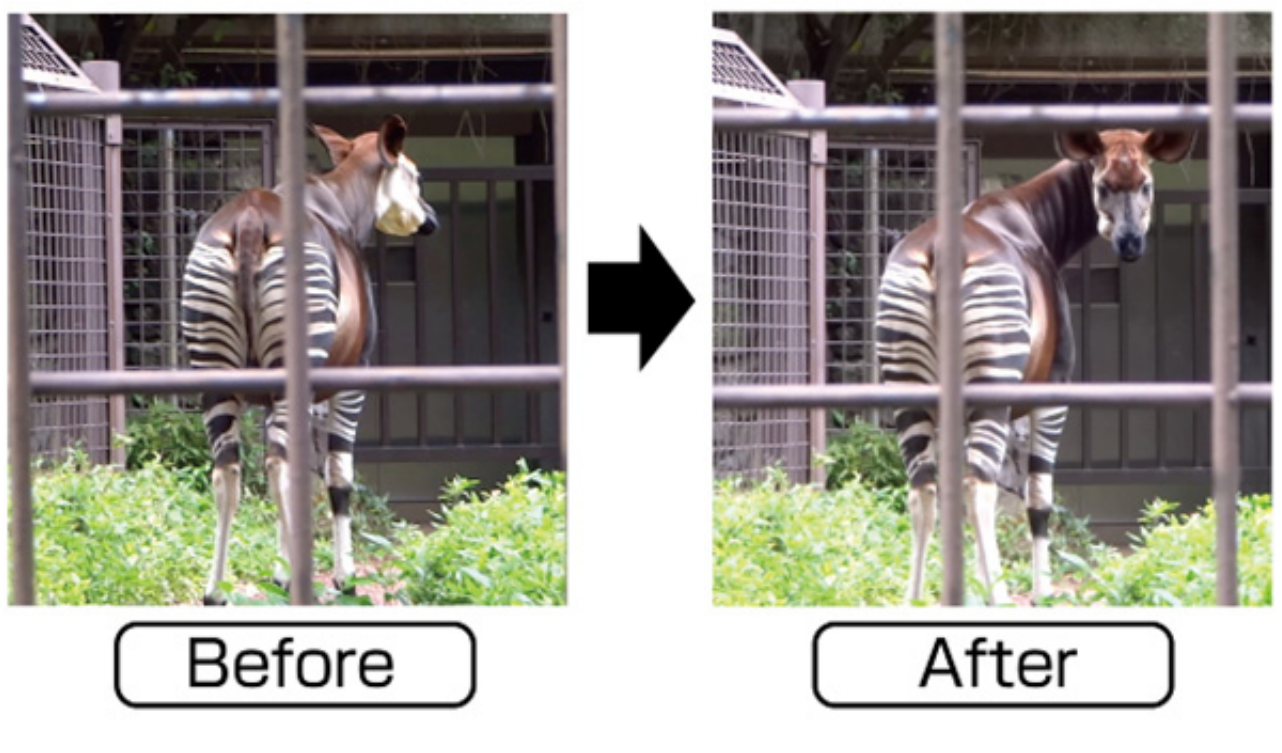

図 1.　AnimalCatcher の目的．動物の多彩な行動を意図的に引き出して撮影する．

Figure 1. The goal of the AnimalCatcher. The system can capture various reactions of animals.

そこで本研究では，撮影中に動物の注意を惹く多様な音を指向性スピーカーで照射することで，意図的に動物のリアクションを引き出すカメラ「AnimalCatcher」を提案する（図 1）．

## 2. AnimalCatcher

AnimalCatcher の主要なコンセプトは以下の 3 点である．

1. 動物の多様なリアクションを引き出す．
2. カメラの画角内で作用する．
3. 従来の撮影スタイルを踏襲する．

第一点は，動物の多様なリアクションを引き出すことである．動物は基本的にカメラを向けられてもリアクションを起こさず，言語で指示を与えることもできないため，一般に良い写真／動画を撮影するためには，根気と運が必要である．そこで，本研究では，動物の写真／動画を撮影する際にその注意を惹く「音」を出力することとした．音の種類を調整することで，動物毎に異なるリアクションを引き出せる可能性がある．

第二点は，音の効果がカメラの画角内で作用することである．単純にスピーカーから音を出力するだけでは音が拡散してしまうため，周囲の迷惑になり，対象の動物が少し

†1 公立はこだて未来大学
Future University Hakodate
†2 津田塾大学
Tsuda College

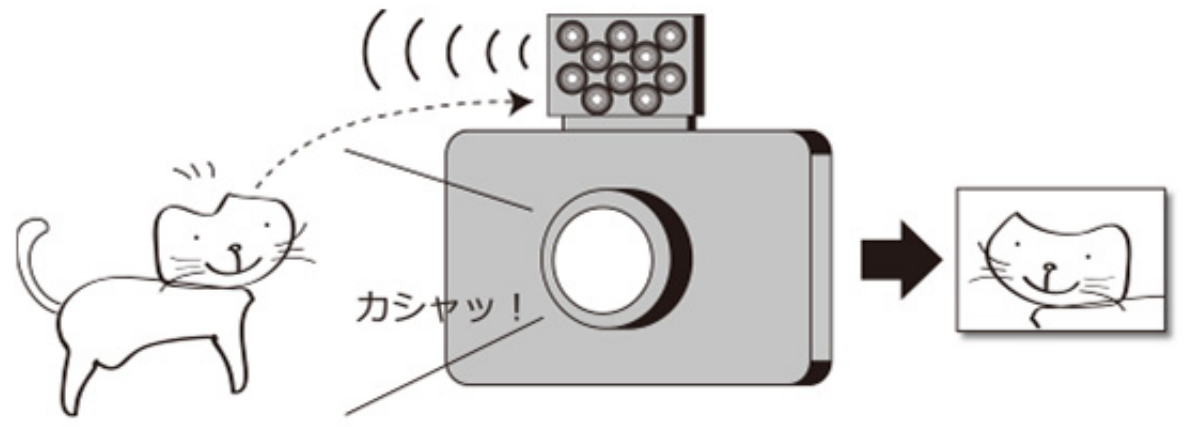

図 2. AnimalCatcher の基本コンセプト．カメラと指向性スピーカーを一体化し，動物の注意を惹きつけつつ写真／動画を撮影する．

Figure 2. The basic concept of the AnimalCatcher. Capturing animals while attracting their attentions by a directional speaker.

離れただけで効果が薄くなってしまう．公園や動物園のような環境で，一般的なカメラで動物を撮影する際の中距離（数 m 以上）で十分効果が発揮できることが望ましい．

この二つの事項を満たすために，我々は超音波素子を用いた指向性スピーカーを利用することにした．一般に指向性スピーカーは直進性に優れる超音波を搬送波として，周波数変調した可聴音を送信することで，数 m～数十 m 程度離れた場所まで直進的に「音」を伝搬できる．この指向性スピーカーでさまざまな「音」を照射することで，カメラの画角の範囲に限定して，動物のリアクションを引き出すことができると考えた（図 2）．

第三点は，カメラを手持ちして写真／動画を撮るという，従来通りの撮影スタイルを踏襲することである．こうした撮影方法は多くの人が慣れ親しんでおり，構図の自由度も高いため，従来の操作体系に倣ったシステムを構築することにする．さらに，動物の表情を捉えるために，高画質な中望遠レンズを備えることが望ましい．

そこで，我々は上述した指向性スピーカーと，WiFi 経由で制御可能なズームカメラ，及びスマートフォンを独自の筐体に組み込んで利用することにした．スマートフォンでは，ヘッドフォン端子を介して指向性スピーカーを制御し，WiFi を介してカメラを制御する．これにより，撮影中の任意のタイミングで動物の耳元に「音」を照射し，高画質の写真／動画でそのリアクションを捉えることができる．

## 3. 実装

AnimalCatcher のプロトタイプの外観を図 3 に示す．プロトタイプは，大きく分けて指向性スピーカーと制御ボード，WiFi カメラ，スマートフォンとこれらを固定する筐体から構成される（図 4）．

プロトタイプはストロボ付きの一眼レフカメラのような外観になっており，下部（幅 140mm×高さ 85mm×奥行 70mm）正面に指向性スピーカーを，背面にスマートフォンを備える．また，上部（幅 70mm×高さ 70mm×奥行 60mm）正面に WiFi カメラを備える．筐体の重さは，全体で約 800g 程度である．決して小型／軽量ではないが，一眼レフのように両手で扱う前提であれば，成人であれば十分保持／撮影ができるよう配慮して設計した．

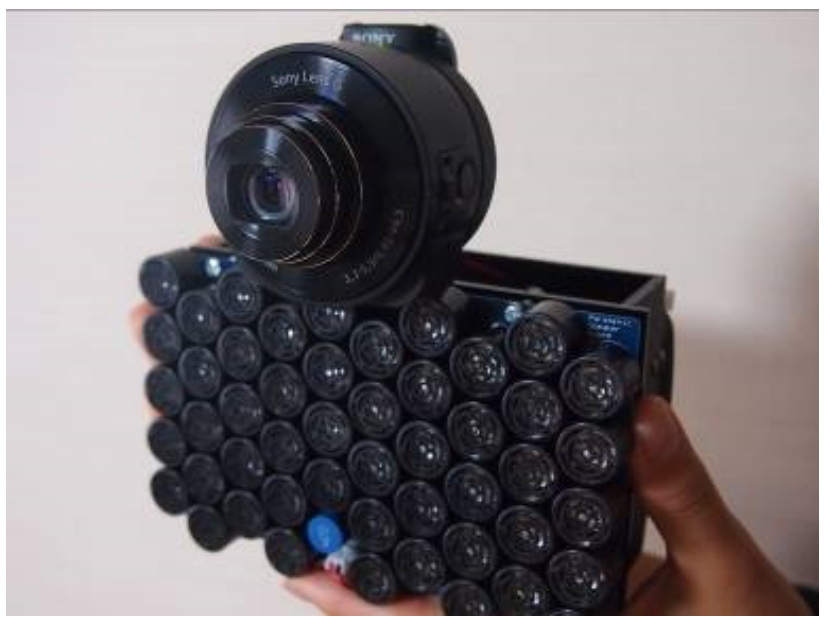

図 3. AnimalCatcher のプロトタイプ外観

Figure 3. The AnimalCatcher prototype.

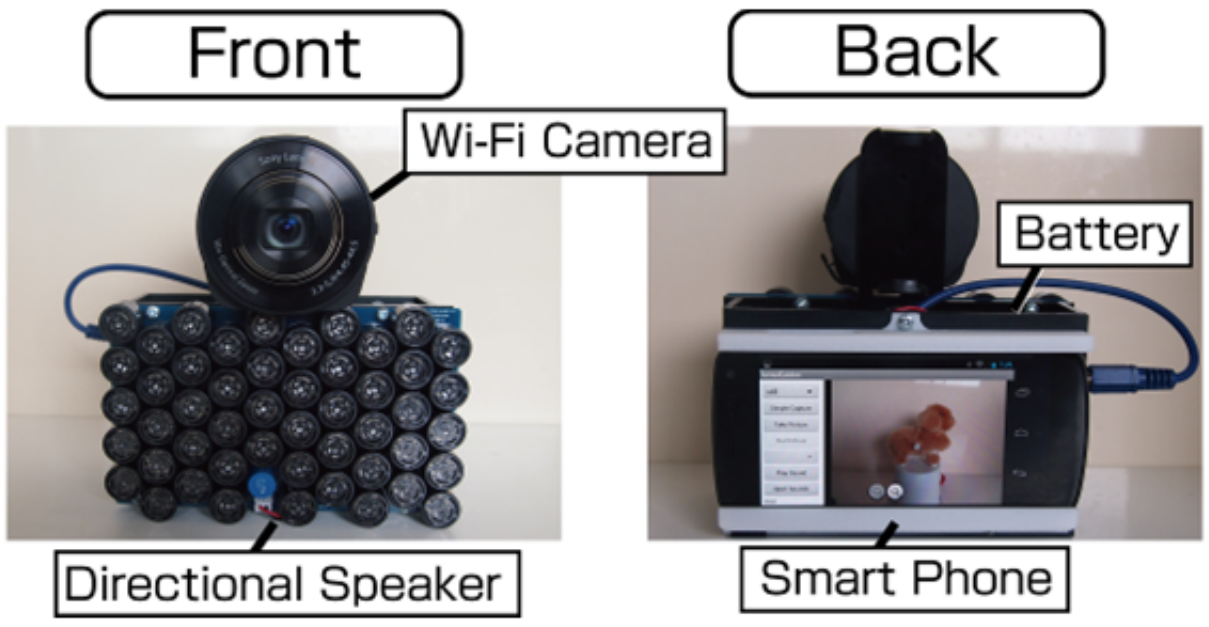

図 4. AnimalCatcher のシステム構成

Figure 4. The system architecture of the AnimalCatcher.

指向性スピーカーとしては，スイッチサイエンスの「超指向性超音波スピーカーキット」 を利用した．これは，秋月電子通商等の同クラス製品と比べて超音波素子と変調回路が改良されており，比較的低価格ながら音量／音質に優れた指向性スピーカーである．指向性スピーカーは制御ボードと接続され，独立した 12V バッテリーで動作する．

WiFi カメラとしては，ソニーのレンズスタイルカメラ DSC-QX10 を利用した．これはその名の通り液晶ファインダー等を備えないレンズ型のカメラであり，WiFi 経由でスマートフォンと連携し，ファインダーやシャッターとして利用する前提で設計されている．なお，DSC-QX10 は内蔵バッテリーで動作する．

スマートフォンとしては，Galaxy Nexus（Android OS 4.2.2）を利用した．スマートフォンのヘッドフォン端子と指向性スピーカーの制御ボードをステレオミニプラグで接続し，DSC-QX10 とは WiFi 経由で接続する．

これらのデバイスを，レーザーカッターで切断したアクリル板と 3D プリンタで出力した樹脂パーツを用いた筐体に固定した．指向性スピーカーと制御ボードは，正面のアクリル板の表／裏にねじ止めし，裏にはさらに電池ケース（単 3×8 本）を樹脂パーツを用いて固定した．DSC-QX10 は，正面のアクリル板の上部に純正のクリップ に合わせた切込みを入れて固定した．Galaxy Nexus は，上下から挟み込むような樹脂パーツを作成し，背面のアクリル板にねじ

止めして固定した．なお，樹脂パーツの固定位置は上下に可変可能であり，ある程度異なるサイズのスマートフォンでも利用可能である．

ソフトウェアは，Android OS 上で Java を用いて実装し，大きく分けてカメラ制御機能と音再生機能を備える．カメラ制御機能は，ライブビュー，静止画／動画撮影，ズーム制御といった基本的な撮影機能を，ソニーの Camera Remote API を用いて実装した．音再生機能は，Android 上のストレージから任意のフォルダ内の音源を一括してロードし，コンボボックスから素早く選択できるよう設計した．なお，撮影方式に応じて，異なる音再生方法を用意した．静止画撮影の場合は，撮影ボタンと音再生ボタンを一体化した．すなわち，撮影ボタンを押し込む間音がループ再生され，離した瞬間に写真を撮影することで，リアクションが起きた瞬間にシャッターを切れるよう配慮した．動画撮影の場合は，撮影途中でリアクションを捉えるため，撮影ボタンと音再生ボタンを分離した．すなわち，動画撮影開始後，音再生ボタンを押し込むと音がループ再生され，離した瞬間に音が停止する仕様とした．

## 4. 運用

本章では，AnimalCatcher の運用として，動物園での撮影事例とその効果について述べる．

### 4.1 目的

本章の主目的は，AnimalCatcher を動物園のさまざまな動物に試用して，その効果（≒動物のリアクションを引き出せるか）を検証することである．さらに，周囲の観客への影響や，デバイス自体の印象についても併せて検証する．

### 4.2 手法

撮影は，2014 年 8 月 25 日・ 29 日の 2 回に分けて東京都内の動物園で行った．この動物園は，多彩な動物が比較的小規模なスペースにまとまっており，都心からアクセスがよい点などを考慮して選択した．1 回目の撮影（予備撮影）では本システムの効果の予備的な検証を行い，2 回目の撮影（本撮影）では実際に本システムを用いて撮影を行った．撮影は 2 人体制で行い，AnimalCatcher を用いて動物を撮影する「撮影者（20 代女性）」と，撮影者と動物，周囲の人々の様子を第三者視点で記録する「監督者（30 代男性）」から構成される．撮影者は予備撮影時にはじめて AnimalCatcher を利用した．各撮影においては，撮影者は AnimalCatcher からの出力音をモニターするため，オーディオ分配器をスマートフォンのヘッドフォン端子に取り付け，ヘッドフォンを装着した．

なお，今回 AnimalCatcher を用いた撮影は全て動画で行った．動画撮影の開始後，撮影者自身が声でキューを出し，直後に指向性スピーカーで音を照射してそのリアクションを記録した．よって，論文上の静止画は，全て動画から切り出したものである．

#### 4.2.1 サウンド設計:

今回，動物に照射する音については，予備撮影では「ホワイトノイズ」を，本撮影では「ホワイトノイズ」「同種の鳴き声」の 2 種類を利用した．

ホワイトノイズは，全ての周波数帯の音を含むため，多くの動物に効果が出ると考え，リファレンス音として用意した．「同種の鳴き声」は，動物が音によって異なるリアクションを引き起こす可能性を調査するために利用した．鳴き声データは，市販の効果音素材集[a]から抽出した．なお，同種の鳴き声は，予備撮影でホワイトノイズの効果が確認でき，素材集に近似種の鳴き声が含まれていた動物約 10 種類分を用意した．なお，音に対する慣れを低減するため，ホワイトノイズと動物の声を用いた撮影の間には，1 時間程度間を空けた．

### 4.3 結果

ここでは，各音源による効果や，周囲の観客への影響等について，具体的な事例を交えながら報告する．

#### 4.3.1 ホワイトノイズの効果

本撮影では，動物園の約 50 種類の動物に対して AnimalCatcher を用いて撮影を試みた．まず，ホワイトノイズを用いた場合の効果一覧を表 1 に示す．表からは，動物が休憩中等の理由で撮影ができなかった動物は省いている．全体的な傾向として，哺乳類，鳥類，爬虫類毎に大きく効果が異なった．

まず，哺乳類については，22 種類中 17 種類の動物が明確なリアクションを行い，うち 14 種類は「派手さ／面白さ」といった観点で，撮影者が魅力的と感じるものだった．次節にて詳しく紹介するが，撮影者に限らず，周囲の観客が動物のリアクションに反応して喜ぶ事例も多数見られた．図 5 に，哺乳類動物のリアクションの一例を示す．

同じホワイトノイズを照射した場合でも，効果は動物によりまちまちであった．ニホンザルやウシのようにこちらを振り向いて凝視するものが多かった（図 5 の 1，2）が，アビシニアコロブスやシマウマのようにこちらを凝視しながら興味深そうに近づいてくるものもいた（図 5 の 3 他）．逆に効果が乏しかった動物としては，アジアゾウやカバが挙げられる．特にアジアゾウは全く効果が見られなかった．なお，これらと外観的に近い印象のあるサイは明確なリアクション（ゆっくりとカメラの方を振り向いて凝視する）が観察できた（図 5 の 4).

キリンやホッキョクグマも本撮影時はリアクションがあまり見られなかったが，これは撮影時の距離が 20～30m 離れていたことに起因すると思われる．実際に，予備撮影ではキリンには明確に効果が出ており（首を起こしてカメラの方を振り向く），同じクマ科でもツキノワグマには効果が出ていた（立つような仕草をする）．

---

a 日本コロムビア 効果音セレクション(2)「動物・鳥・蛙・虫」
http://www.amazon.co.jp/dp/B00CRKLG68

表 1. AnimalCatcher でホワイトノイズを照射した際のリアクション効果.
◎: 明確なリアクション&魅力的な写真が取れたb
○: 明確なリアクションを引き出した.
△: 微かにリアクションした.
×: 全く反応しなかった.
Table 1. The effects of the AnimalCatcher using white noise

| 名前 | 大分類 | 小分類 | 効果 |
|---|---|---|---|
| ニホンザル | 哺乳類 | サル目 | ◎ |
| エリマキキツネザル | 哺乳類 | サル目 | ◎ |
| アビシニアコロブス | 哺乳類 | サル目 | ◎ |
| ウシ | 哺乳類 | ウシ科 | ◎ |
| バーバリーシープ | 哺乳類 | ウシ科 | ◎ |
| ヤギ | 哺乳類 | ウシ科 | ◎ |
| 子羊 | 哺乳類 | ウシ科 | ◎ |
| シマウマ | 哺乳類 | ウマ目 | ◎ |
| ウマ | 哺乳類 | ウマ目 | ○ |
| プレーリードッグ | 哺乳類 | ネズミ目 | ◎ |
| カピバラ | 哺乳類 | ネズミ目 | ○ |
| オカピ | 哺乳類 | キリン科 | ◎ |
| キリン | 哺乳類 | キリン科 | △ |
| ゼニガタアザラシ&アシカ | 哺乳類 | 食肉目 | ◎ |
| カンガルー | 哺乳類 | カンガルー目 | ◎ |
| タテガミオオカミ | 哺乳類 | イヌ科 | ◎ |
| サイ | 哺乳類 | サイ科 | ◎ |
| コビトカバ | 哺乳類 | カバ科 | △ |
| クマ（3 種類） | 哺乳類 | クマ科 | △ |
| アジアゾウ | 哺乳類 | ゾウ目 | × |
| シロフクロウ | 鳥類 | フクロウ目 | ◎ |
| ワシミミズク | 鳥類 | フクロウ目 | ◎ |
| バラワンコクジャク | 鳥類 | キジ目 | △ |
| ニワトリ | 鳥類 | キジ目 | △ |
| ダルマワシ | 鳥類 | コウノトリ目 | ◎ |
| タンチョウヅル | 鳥類 | ツル目 | ◎ |
| ペンギン | 鳥類 | ペンギン目 | ◎ |
| オウギバト | 鳥類 | ハト目 | ○ |
| フラミンゴ | 鳥類 | フラミンゴ目 | △ |
| ハシビロコウ | 鳥類 | ペリカン目 | ○ |
| コンドル | 鳥類 | タカ目 | △ |
| ワニ（2 種類） | 爬虫類 | ワニ目 | × |
| カメ（3 種類） | 爬虫類 | カメ目 | × |
| オオトカゲ | 爬虫類 | 有鱗目 | × |

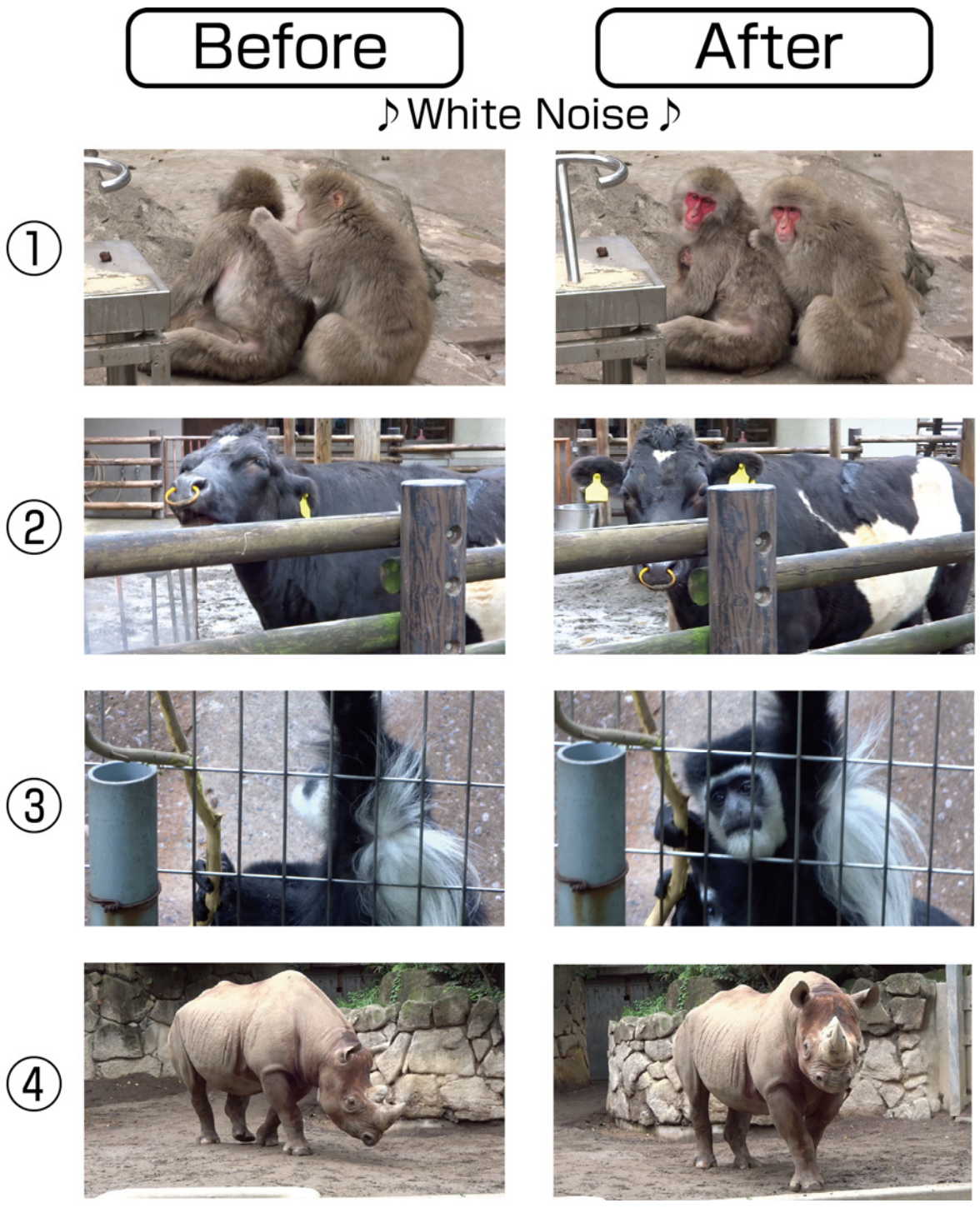

図 5 哺乳類動物の撮影事例.
1 ニホンザル: 二匹同時にこちらを振り向く. 2 ウシ: 耳をピンとさせた後こちらを振り向く. 3 アビシニアコロブス: 木登り中にこちらを振り向き，興味深そうに近づく. 4 サイ: 悠然と歩いていたが，耳を動かして音源を探った後，こちらに振り向く.
Figure 5 Example pictures of mammals. 1 Japanese macaque, 2 Cattle, 3 Abyssinian Black-and-white Colobus, 4 Rhinoceros

次に，鳥類については，11 種類中 7 種類が明確なリアクションを行い，うち 5 種類は「派手さ／面白さ」といった観点で，撮影者が魅力的と感じるものだった．図 6 に，鳥類のリアクションの一例を示す．鳥類で特徴的なのは，目が顔の側面についている動物が多いため，カメラ目線を送っているにも関わらず撮影者からは分かりにくいものがいることである（図 6 の 1)．一方，フクロウ系は目が顔の正面についているため，まっすぐカメラ目線で撮影することができる（図 6 の 2).

最後に，爬虫類については，ワニ／カメ／トカゲといった 7 種類の動物で検証したが，ほぼ全く効果が見られなかった．

### 4.3.2 動物の声の効果

まず，同種の動物の声は，ホワイトノイズの効果があっ

b ここで，◎／○の差は主にリアクションの派手さ／面白さに，○／△／×の差は純粋にリアクションの大きさに起因する．

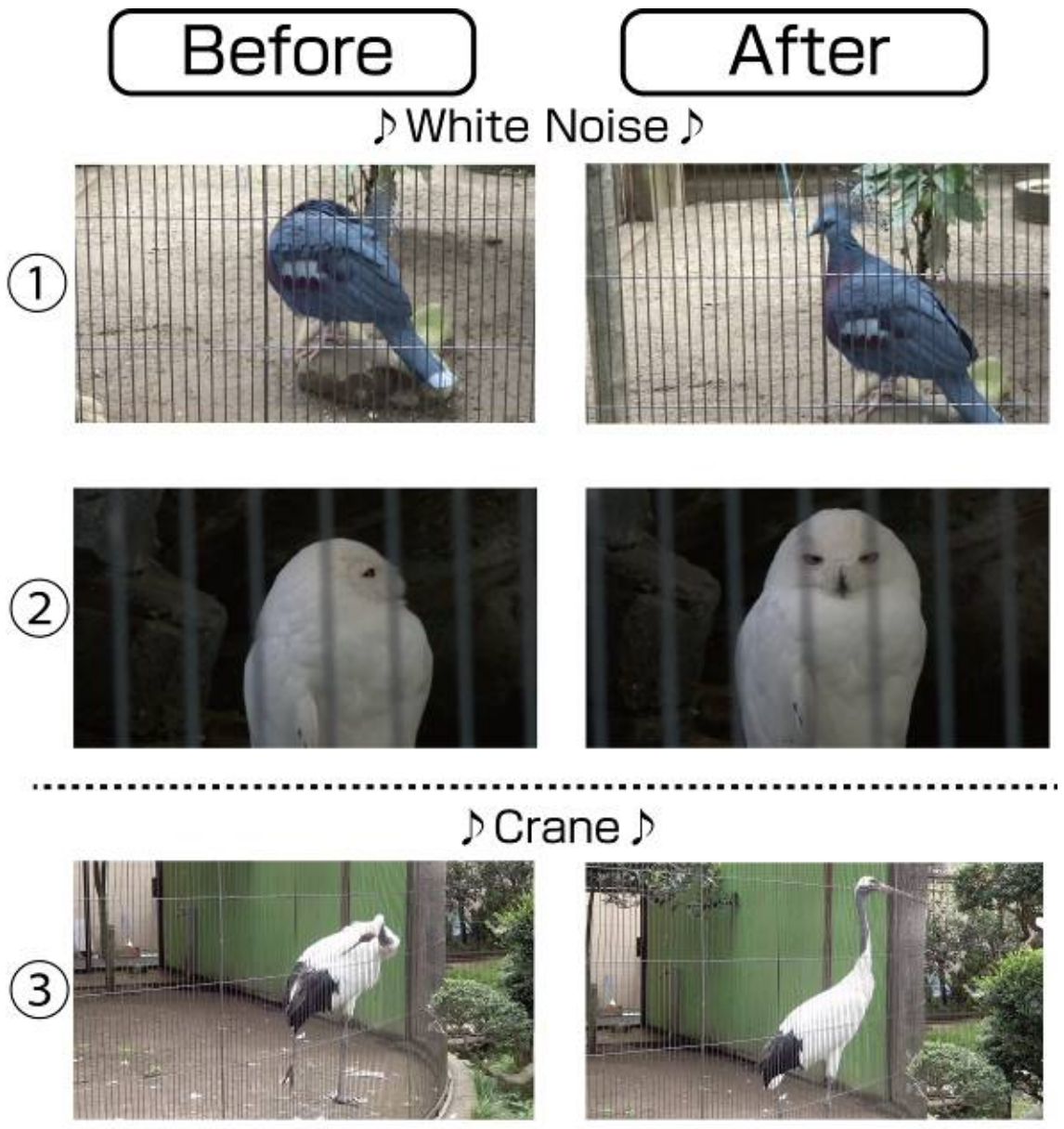

図 6 鳥類の撮影事例．1 オオギバト: 体を起こして視線を送るが，目が顔の横にあるためカメラ目線に見えない．2 シロフクロウ: 素早く振り向き凝視する．3 タンチョウヅル: 体を起こして視線を送る．ホワイトノイズ／ツルの鳴き声に同じように反応した．

Figure 6 Example pictures of birds. 1 Victoria crowned pigeon. 2 Snowy owl, 3 Red-crowned crane.

た動物のうち，適切な音源を用意できた哺乳類 3 種類，鳥類 5 種類に対して検証した．その結果，明確な効果があったのは，鳥類のタンチョウヅルのみだった．タンチョウヅルは，ツルの鳴き声に対して，ホワイトノイズの場合とほとんど同じリアクション（首を挙げてゆっくり視線を送る）を見せた（図 6 の 3)．シマウマについて，ホワイトノイズと同種の鳴き声を適用した様子を図 7 に示す．ホワイトノイズに対しては，耳を立てて顔をあげた後、音のする方向に近づいてきた．一方，草を食べている途中ではあったが，ウマの声にはほぼ全く反応しなかった．

### 4.4 周囲の観客への影響

ここでは，AnimalCatcher が周辺の観客に与えた影響について，実例を詳しく紹介する（図 8)．なお，下記項目は図 8 中の番号に対応している．

1. アシカやアザラシが突然動き出したため，弟を抱えながら身を乗り出していた姉がのけぞり，「うわびっくりした！」と声を上げた．アシカの突然の行動に驚いたように，「なんでなんで？」と続けた．
2. ホッキョクグマの気をひこうと，老人が手拍子をしている．その真横で AnimalCatcher を利用しているが，デバイスの音は全く聞こえていないようだ．
3. オカピを美しく振り向かせられた（図 1 に相当）ので，思わぬ珍事にとなりの婦人が「すごーい！」と口に手

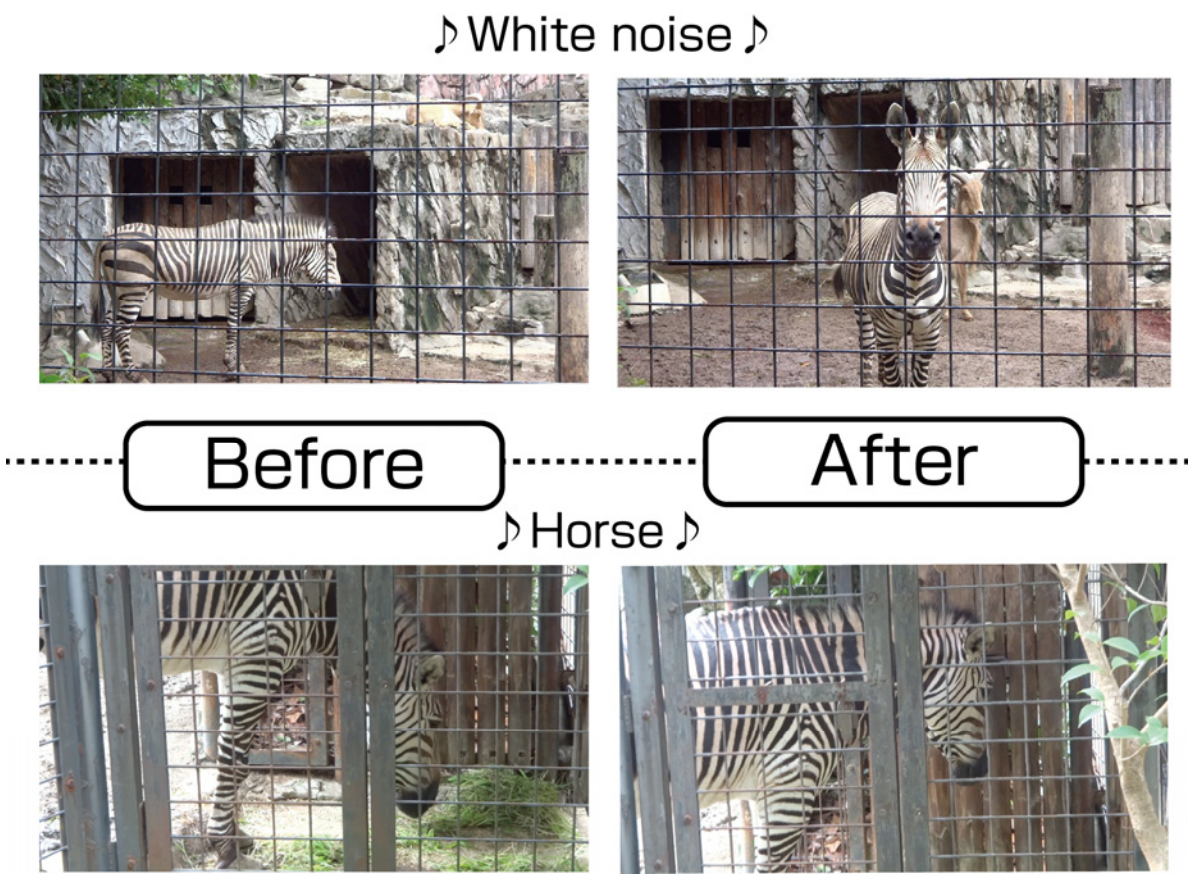

図 7 シマウマの撮影事例．上段: ホワイトノイズに対して，耳を立てて顔をあげた後、音のする方向に近づいてくる．下段: 馬の声に対して全く反応せず草を食べ続ける．

Figure 7 Example pictures of a zebra. Upper: With white noise. Lower: With a horse’s voice.

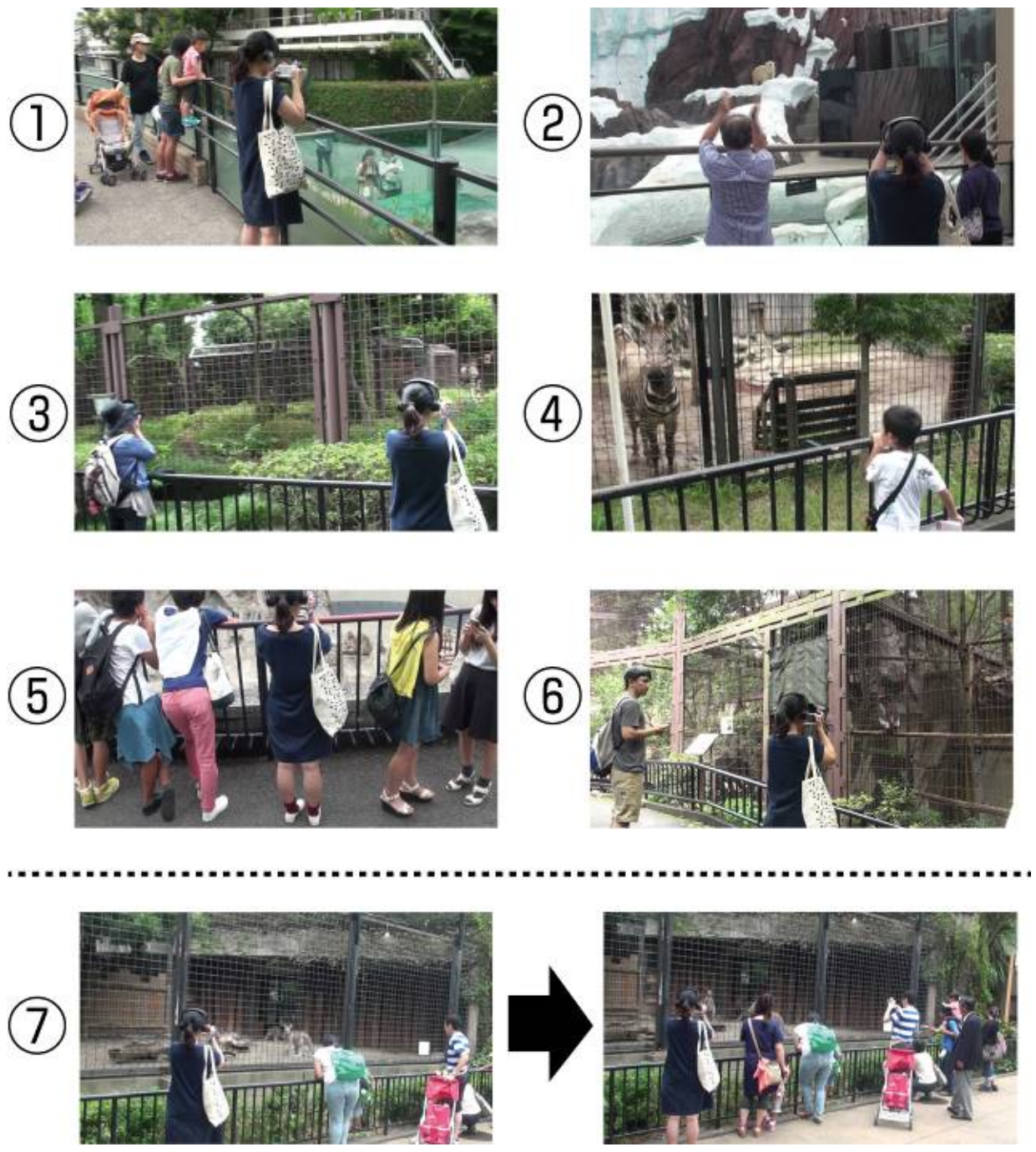

図 8 周囲の人々への影響事例．白鞄の女性は撮影者．1 アシカの反応に驚く姉弟，2.ホッキョグマの気を惹く老人，3 振り向くオカピに喜ぶ女性，4 近づくシマウマに喜ぶ子供，5 振り向くニホンザルに喜ぶカップル，6 撮影者を気にする男性，7 カンガルーの反応に喜んで集まった観衆．

Figure 8 Examples of effects on surrounding people.

を当てて興奮している．

4. シマウマをカメラの近くに近づかせることに成功した（図 7 上段に相当）ので，子どもが興奮して「なんで会えたの？（※近づいてきたの？）」と叫んでいた．

5. ニホンザルを 2 匹一緒に振り向かせることに成功した（図 5 の 1 に相当）ので，左の二人連れが，「二人（※二匹？）で見つめられるの，すばらしい」と喜んでいた．デバイスの音には全く気付いていないようだ．
6. ダルマワシの写真を撮っていた男性が撮影者の持つデバイスに気づいて凝視するが，何が起きているかは理解できなかったようだ．
7. AnimalCatcher を利用した所，カンガルーの集団が同時に大きく動き出したので，周囲の観客が集まり，一斉にシャッターを切った．肩車の親子は「あー，動いてる！」と叫んだ．

### 4.5 デバイス自体の印象

ここでは，主に本撮影における，撮影者と監督者によるデバイス自体の印象についてまとめる．

- 2 時間ほどで WiFi カメラのバッテリーが切れてしまったため，本撮影の途中で充電した．一方，スマートフォンと指向性スピーカーのバッテリーは 3 時間（休憩除く）の撮影中一度も電源を切らなかったが，最後まで持続した．
- デバイスの使い勝手は，起動した状態を常に保てば特にストレスは感じなかった．WiFi カメラの初期接続に少し時間がかかる（約 5～10 秒程度）ので，細かく電源を切りたい場合は多少不便かもしれない．
- カメラを構えるときに手が少しスピーカ素子を覆うことが多いので, 取っ手のようなものはあるとよさそう．
- 撮影操作／音声再生操作について失敗はなかったが，右利き／左利きに合わせて，ボタンの配置を左右に調整できるとよりよさそう．
- デバイス自体から音が出力されているかを確認したい．今回はヘッドフォンと分岐ジャックを利用したが，システム側でも何らかのサポートがあるとよい．
- ヘッドフォンは別に邪魔ではなかったが，周囲に「この人本格的に音を扱ってる！」という印象を持たれて，目立ってしまうように感じた．
- ズーム操作は，タッチパネルの応答がやや遅かったので，カメラ本体のハードウェアボタンで行うことが多かった．

### 4.6 運用のまとめ

ここでは，動物園での運用と撮影事例から得た知見をまとめる．まず，AnimalCatcher の動物のリアクションを引き出す効果については次のようにまとめられる．

- 試行回数は 1～2 回だが, 哺乳類と鳥類に対しては概ね大きな効果があり，爬虫類には全く効果がなかった．
- 有効射程距離は概ね 10m 程度だと思われる．20m 程度離れた状況では，効果が期待できる動物も反応しなかった．
- 音の種類については，ホワイトノイズが予想以上に効果的であった．同種の声は全体的には効果が低かったが，タンチョウヅルに対しては効果があった．今後は音と効果の関係を精査したい．

次に，周囲の鑑賞者への影響については，以下のようにまとめられる．

- 照射する音にはほとんど気づかれなかった. 図 8 の 2，5 のように両隣に密接して別の観客がいた場合でも全く気付かれなかった．
- 図 8 の 3，4，5，7 のように，周りの客も予期せぬ動物の動作を見て，一緒に喜ぶケースが散見された．

最後に，デバイス自体の使い勝手については，以下のようにまとめられる．

- バッテリーの連続稼働時間は WiFi カメラがボトルネックで，2 時間程度である．こまめに電源を切れば改善するが，WiFi カメラの初期接続にはやや時間がかかる欠点もある．
- デバイスの音は屋外などの開けた環境では撮影者自身にもほとんど聞こえないため，システム側で音声レベルを表示するなどのサポートを検討する．
- デバイスの重量については, 直接不満は出なかったが，持ち手やストラップ端子を付けるなどして可搬性を改善したい．

## 5. 議論

ここでは,「動物病院での撮影」「獣医からのフィードバック」「動物愛護」の観点から AnimalCatcher について議論する．

### 5.1 動物病院での撮影

動物園での運用とは別に, AnimalCatcher を都内の動物病院に持ち込み，獣医の指導の下で撮影を行った．この動物病院は家庭で飼育される犬や猫等のペットを中心に扱っている．よって，撮影対象は主に犬と猫であり，音源としては,「ホワイトノイズ」と「同種の鳴き声」を用いた．同種の鳴き声は，犬については「甘え、悲鳴、威嚇、けんか、群れ、遠吠え」といった複数の感情を，猫については特に規定のない一種類の声のみを用意した．以下，箇条書きで結果を示す．

- ホワイトノイズを照射すると，犬／猫共に持続的にカメラの方を向いたり，耳を回して音源方向を探るような動作を引き出す効果が高確率で確認された．
- 犬に犬の声や，猫に猫の声を照射した場合も，持続的にカメラの方を向く効果が確認された．一方，声が間歇的に発音され平均的な音圧が弱いせいか，効果はホワイトノイズに比べてやや弱いように思われた．
- 犬に様々な感情の犬の声を連続で聞かせたが，慣れの効果が大きかったせいか，順番が後の音声ほど目立った効果が確認できなくなる傾向が見られた．

犬／猫に適用したいくつかの具体例を図 8 に示す．

### 5.2 獣医からのフィードバック

ここでは，動物園での運用と動物病院での撮影の結果と，獣医からのフィードバックを合わせて議論する．

まず，獣医に AnimalCatcher の基礎的な原理（遠隔地から多様な音を照射する）を説明した所，「犬は目の前の音よりも見えないところの音に興味をもち反応する．たとえば，暴れるイヌの体重を測る時などは，わざと見えない所で机を叩くなどすると動くのをやめて，静かに音の方向を探ろ

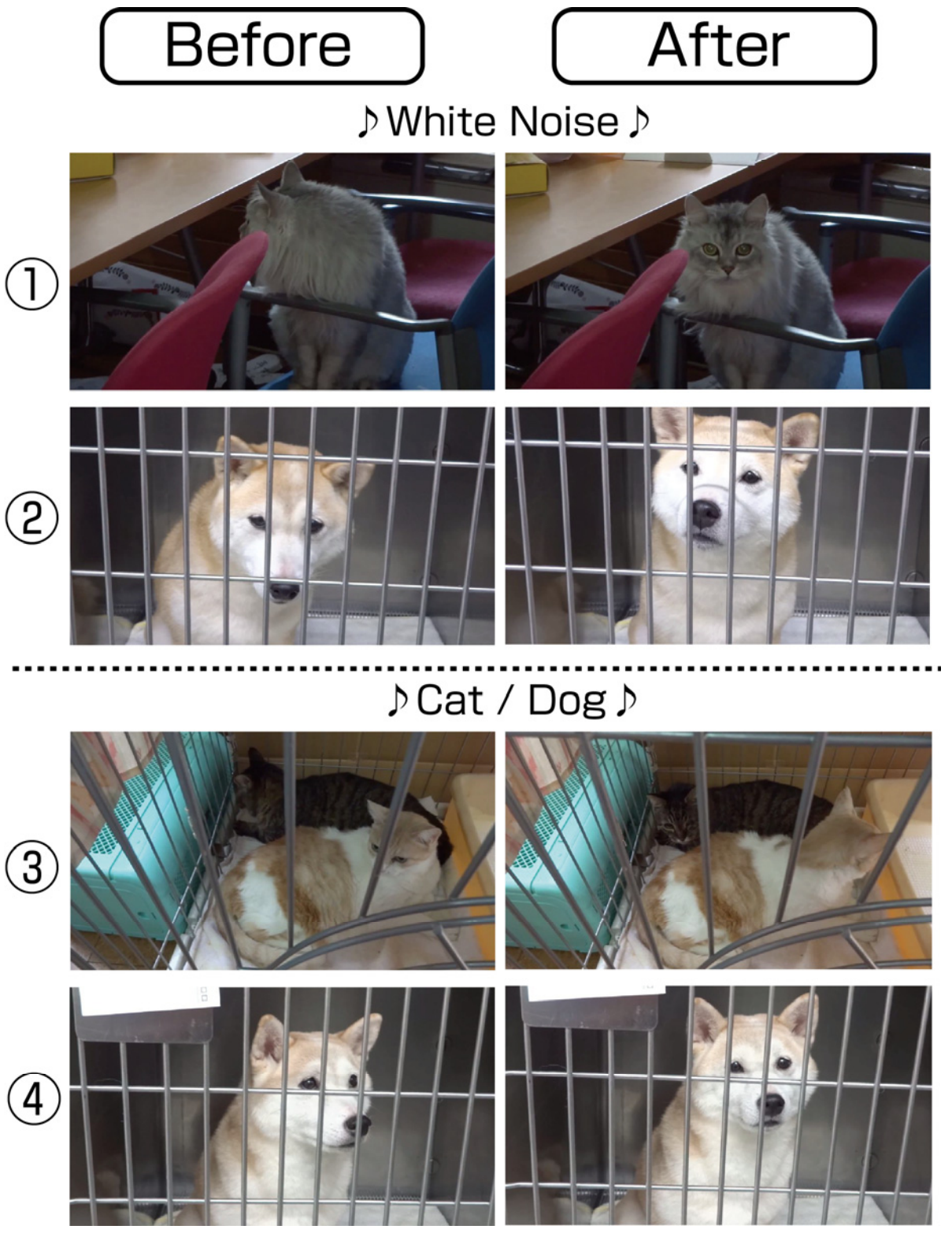

図 8 動物病院での犬／猫の撮影事例．1. ホワイトノイズに猫が素早く振り向きカメラを凝視する．2. ホワイトノイズに犬が瞬きをした後、鼻をくんくんさせる．3. 猫の声に反応し，2 匹の猫が顔を反転させ，顔や耳を動かす．4. （2 と同一の）犬が，犬の声に反応し，カメラ目線でじっと見つめる．

Figure 8 Example pictures of dogs/cats in an animal hospital.

うとする．その際に体重を測るというノウハウがある」という事例が紹介された．この点で，AnimalCatcher の，「見えない音を当てる」，というアプローチは動物の興味を引く上で一定の合理性があると思われた．

次に，今回「同種の動物」として選んだ音源は，基本的に音源 CD から選択したものである．たとえば，犬に対しては「甘え」「怒り」といったいくつかの音声を照射したが，それぞれの犬の種類までは考慮していない．よって，ブルドックに対してトイプードルの声を聞かせる，のような状況であった．我々はこの音源選択の妥当性にやや不安を持っていたが，獣医によると「少なくとも犬については，種類は違っても言語的にはほぼ同一なので，このように動物の種類と音源の種類が完全に一致していなくても問題ないのでは」とのコメントが得られた．

また，鳴き声などを照射する際の注意点として，「犬には『激怒症候群』という障害（遺伝病で急に猛烈に怒る）を持つ個体がいるため，念のため威嚇とか喧嘩の声は聞かせない方がよい．」という指導を受けたケースがあった．他の動物においても，音源を選択する際にこうした種固有の事情や障害の可能性などに配慮する必要があると考える．

さらに，動物園の動物に対するコメントとして，「動物園の動物は，野生のもの連れてくるケースは最近では少ない．動物愛護の観点から．何らかの形で野生で生きられなくなったり，園内で繁殖したものが大半である．なので，本当の野生動物よりも音に関する感覚・関心は鈍い可能性がある．」との意見が得られた．これは，本物の野生動物に対しては今回の調査とは異なる効果が出る可能性を示唆している．

また，ペットに対して利用した場合に「慣れ」による性能低下が大きいように感じたため，この点について獣医に確認したところ，「ペットは人間界に慣れているので，いろいろな刺激にすぐ慣れるし鈍感になる．たとえば，動物病院でも，生まれたばかりに子猫を，わざと人の往来のあるところで飼育して，人間界の刺激に慣らせることがある．」とのコメントが得られた．人間の生活圏に近いほど，連続使用時の効果が減少する可能性があるため，今後はこうした「慣れ」の問題についても考慮した設計を進めたい．

### 5.3 動物愛護

運用の結果，AnimalCatcher はかなり効果的に動物のリアクションを引き出すことが確認できた．一方で，AnimalCatcher を実際に社会に普及させるフェーズを考えると，特に連続使用時に，前述した「動物が刺激に対して慣れてしまう」可能性に加えて，「動物にストレスを与える」可能性も当然考慮する必要がある．ここでは，こうした動物愛護的な観点から議論する．

まず，AnimalCatcher に関連しうる法律上の規定としては，動物愛護法が愛護動物（ペット）をみだりに傷つけることを禁じている． さらに，法的に問題ない場合でも，様々な宗教的思想・信念に基づく過激な動物愛護団体の批判にさらされる可能性にも留意しなければならない．

AnimalCatcher の利用シーンとして，ユーザが自分の飼育するペットに対し，よりよい写真撮影を行うために常識的な範囲内で利用することには大きな問題はないだろう．一方で動物園などの公共施設において，来場者がそれぞれ AnimalCatcher を所持して無秩序に使用するような状況は，動物のストレスを適正範囲に抑える上では問題がある．これを回避するために公共施設が実施しうる AnimalCatcher にかかわる運用方針には，現状で施行されている「フラッシュ撮影禁止」と「お披露目タイム・餌やりタイム」を基にした 2 種類が考えられる．前者のフラッシュ撮影は，よ

り良い写真撮影のために一般的であり，ほぼ全てのカメラに搭載されている技術であるが，動物への悪影響が指摘されており，動物園では使用が禁止されることが多い．AnimalCatcher も特定の動物への悪影響が心配される場合は，ある音源／あるいはデバイス自体の使用を施設側が禁止すればよい．これが最も簡単な動物愛護対策である．一方で「お披露目タイム」は来場者の鑑賞にさらされる時間を施設側が設定する手法であり，「餌やりタイム」とは，食事という普段見られない動物の生態を鑑賞／写真撮影するために施設側が特別な時間を用意する手法である．いずれも動物へのストレスを施設側が一括して調整／管理することが可能である．AnimalCatcher の使用についても，個々の来場者による無秩序な使用を禁止した上で，施設側がシステムを準備し公式イベントとして適切な時間間隔で「AnimalCatcher タイム」として運用することで，動物に過度なストレスがかからないように調整しつつ，来場者のよりよい鑑賞体験および写真撮影を実現することが可能であろう．

さらに，より社会に馴染みやすい技術とするために，システムの実装方法を改良し，従来の動物撮影により近づける方法も検討している．すなわち，自分の声をマイクで拾って，ホワイトノイズと任意の割合で混合して照射する，といった方式である．この方式は，(1)現状でも音声で動物に呼びかけることは一般的である，(2)声を発している間でしか連続使用できない，といった特徴を持ち，あたかもユーザの呼びかけた声に動物が自然に反応する状態を実現できると考えられるため，一方的に抑圧するような印象を改善できると考える．

なお，指向性スピーカーの音量については，ホワイトノイズを 2m 程度の距離から人間に直接照射して聞いた状態でも，「さーという音が認識できる」程度であり，「うるさくて耐えられない」という程ではない．動物がこの音をどう認識しているかを正確に判断することは難しいが，元々動物園では日々多くの人が訪れて騒がしく，様々な音が混在しているため，ホワイトノイズもそこまで特別な音ではなく，連続使用するとすぐ慣れてしまう程度に害のない音なのではないかと考えている．

## 6. 関連研究

撮影時にコンテンツを提示することで被写体のリアクションを引き出す研究がある．EyeCatcher[3]は，デジタルカメラの上部に小型のディスプレイを装着し，シャッターを切る直前にコンテンツを表示することで，多彩な表情を撮影するシステムである．CheeseCam[1]は，画面上に表示された顔アイコンを見た時に人が無意識的にリアクションを起こすことを報告している．伏見ら[6]は，撮影時に笑い声を音で提示することで，被写体の笑顔を誘発するシステムを提案している．これらのシステムは，人間の被写体を対象としているのに対し，本研究は動物の被写体を対象としている点が異なる．

ねこ猫カメラ[5]は，猫の鳴き声など猫の気を惹きやすい音を再生しながら撮影することで，動物写真の撮影を支援するスマートフォンアプリである．本研究と狙いは共通するが，猫のみを対象としており，スマートフォンのスピーカーを利用するためごく近距離でしか効果を発揮しない．本研究では，さまざまな種類の動物に対して，10m 程度離れてもリアクションを引き出せる点が特徴である．

カメラにさまざまなセンサを搭載し，撮影時のコンテキスト情報を記録するシステムとしては，ContextCam[2]，WillCam[4]等がある．本研究は，動物を撮影するという撮影行為そのものに着目し，動物の行動を意図的に引き出す点が異なる．

## 7. まとめ

本論文では，カメラでの撮影中に動物の行動を誘発する多様な音を指向性スピーカーで照射することで，瞬間的に動物のリアクションを引き出すカメラ「AnimalCatcher」を提案，実装した．さらに，動物園での運用を通して，哺乳類／鳥類の多くの動物のリアクションを引き出せることを確認した．今後は，議論で述べたような獣医などの専門家の知見や動物愛護的な観点を踏まえつつ，適切に社会に導入する方法を検討していきたい．

## 参考文献